\documentclass[conference]{IEEEtran}
\IEEEoverridecommandlockouts
\usepackage{cite}

\usepackage{amsmath,amssymb,amsfonts}
\usepackage{hyperref}
\usepackage{graphicx}
\usepackage{bbding}
\usepackage[misc]{ifsym}
\usepackage{makecell}
\usepackage{booktabs}
\usepackage{subfigure}
\usepackage{tabularx}
\usepackage{amsmath}
\usepackage{multirow}
\usepackage{algorithm}
\usepackage{etoolbox}
\usepackage[justification=justified]{caption}
\usepackage{etoolbox}
\patchcmd{\thebibliography}{\section*{\refname}}{}{}{}
\makeatletter
\patchcmd{\@makecaption}
  {\scshape}
  {}
  {}
  {}
\patchcmd{\@makecaption}
  {\\}
  {.\ }
  {}
  {}
\makeatother
\usepackage{hyperref} 
\hypersetup{
hidelinks,
colorlinks=true,
linkcolor=red,
citecolor=green
}

\usepackage{algorithmic}

\usepackage{graphicx}
\usepackage{textcomp}
\usepackage{xcolor}
\def\BibTeX{{\rm B\kern-.05em{\sc i\kern-.025em b}\kern-.08em
    T\kern-.1667em\lower.7ex\hbox{E}\kern-.125emX}}
\begin{document}

\title{SATBA: An Invisible Backdoor Attack Based on Spatial Attention\\
\thanks{Corresponding author:Xiaowei Xu(xuxw525@ouc.edu.cn)}
}

\author{\IEEEauthorblockN{1\textsuperscript{st} Huasong Zhou}
\IEEEauthorblockA{\textit{College of Computer Science and Technology} \\
\textit{Ocean University of China}\\
Qingdao, China \\
zhouhuasong@stu.ouc.edu.cn}
\and
\IEEEauthorblockN{2\textsuperscript{nd} Xiaowei Xu}
\IEEEauthorblockA{\textit{College of Computer Science and Technology} \\
\textit{Ocean University of China}\\
Qingdao, China \\
xuxw525@ouc.edu.cn}
\and
\IEEEauthorblockN{3\textsuperscript{rd} Xiaodong Wang}
\IEEEauthorblockA{\textit{College of Computer Science and Technology} \\
\textit{Ocean University of China}\\
Qingdao, China \\
wangxiaodong@ouc.edu.cn}
\and
\IEEEauthorblockN{4\textsuperscript{th} Leon Bevan Bullock}
\IEEEauthorblockA{\textit{College of Computer Science and Technology} \\
\textit{Ocean University of China}\\
Qingdao, China \\
leonbevanbullock@ouc.edu.cn}
}

\maketitle

\begin{abstract}
Backdoor attack has emerged as a novel and concerning threat to AI security. These attacks involve the training of Deep Neural Network (DNN) on datasets that contain hidden trigger patterns. Although the poisoned model behaves normally on benign samples, it exhibits abnormal behavior on samples containing the trigger pattern. However, most existing backdoor attacks suffer from two significant drawbacks: their trigger patterns are visible and easy to detect by backdoor defense or even human inspection, and their injection process results in the loss of natural sample features and trigger patterns, thereby reducing the attack success rate and model accuracy. In this paper, we propose a novel backdoor attack named SATBA that overcomes these limitations using spatial attention and an U-net based model. The attack process begins by using spatial attention to extract meaningful data features and generate trigger patterns associated with clean images. Then, an U-shaped model is used to embed these trigger patterns into the original data without causing noticeable feature loss. We evaluate our attack on three prominent image classification DNN across three standard datasets. The results demonstrate that SATBA achieves high attack success rate while maintaining robustness against backdoor defenses. Furthermore, we conduct extensive image similarity experiments to emphasize the stealthiness of our attack strategy. Overall, SATBA presents a promising approach to backdoor attack, addressing the shortcomings of previous methods and showcasing its effectiveness in evading detection and maintaining high attack success rate.

\end{abstract}

\begin{IEEEkeywords}
backdoor attack, deep neural network, spatial attention, U-net
\end{IEEEkeywords}

\section{INTRODUCTION}
Various domains such as facial recognition\cite{hammouche2022gabor}, automated driving\cite{wang2023centernet}, medical diagnosis\cite{nijaguna2023quantum}, etc., have witnessed the impressive performance of Deep Neural Network (DNN) in the last decade. However, they also expose a serious weakness to adversarial attack\cite{goodfellow2014explaining}, which can manipulates the prediction output of DNN by adding small noises to the input samples.

Backdoor attack is a stealthy attack against DNN that has emerged with the advancement of attack techniques. Unlike traditional adversarial attack, it seeks to insert a hidden backdoor into the DNN, which makes the target model behave normally on clean samples but change its output when the hidden backdoor is activated by the attacker's designed input. This allows the attacker to control the DNN.

There are numerous approaches to implement a backdoor attack, such as directly modifying clean dataset and training the model on it\cite{gu2019badnets,li2022untargeted,wang2022invisible} or altering the model weights during the model deployment phase\cite{hong2022handcrafted,chen2021proflip}, but the primary attack typically involve implanting the backdoor during the model training stage.

Despite the improvements in backdoor attack, most existing methods of backdoor attack still face the following major challenges: (1) The trigger is easily identified and removed by backdoor defenses and even by humans because it is conspicuous. (2) The features of images and trigger patterns are usually lost during the attack stage as a result of the modification of the clean image and trigger in the spatial domain.

This paper proposes SATBA, a novel imperceptible backdoor attack on DNN that exploits spatial attention. Our attack consists of three steps: (1) extracting image features using a conventional algorithm, (2) acquiring the spatial attention weights of the victim model on clean images and generating trigger patterns based on them, and (3) employing an U-shaped convolutional neural network to embed the trigger in benign images, thereby initiating an attack on target model's training process. To validate the effectiveness and reliability of our attack, we conducted extensive experiments on benchmark datasets and DNNs. The results demonstrate a high Attack Success Rate (ASR), while ensuring a great Clean Data Accuracy (CDA). Additionally, SATBA exhibits a low anomaly index and operates with a high level of stealthiness.

We summarize the main contributions of this paper as follows:
\begin{itemize}
\item This paper presents the SATBA attack, which is the first attempt to use spatial attention to create trigger pattern and install backdoor in DNNs.
\item An U-net based network is designed to insert trigger patterns into clean images with minimal feature loss. The network preserves feature of both clean images and triggers throughout the injection process.
\item Extensive experiments have shown that SATBA has comparable attack success rate to several common backdoor attacks, while outperforming them in terms of robustness and stealthiness, demonstrating the versatility and effectiveness of our approach.
\end{itemize}

The rest of this paper is organized as follows. In Section II, we briefly review the related works of backdoor attack, backdoor defense, U-net and spatial attention. In Section III, we present the details of our proposed attack method. Our experimental results are reported and analyzed in Section IV. Section V is reserved for a discussion regarding the implications and the future works of our research. Finally, we conclude this paper in Section VI.

\section{RELATED WORKS}
In this section, we provide a review of related works on backdoor attack and defense, followed by explanations of U-net and spatial attention techniques.
\subsection{Backdoor Attack}
Backdoor attack on DNNs was first introduced by Gu et al.\cite{gu2019badnets}, who inserted a small patch into clean images and used the altered images to train the target model. This attack, also known as Badnets or Patched attack, could manipulate the behavior of the trained model when it is exposed to specific images with the trigger pattern (a small patch). Based on it, Chen et al.\cite{chen2017targeted} suggested a different strategy for trigger generation and injection, using a watermarked image as a trigger and optimizing the weights between benign and trigger images. This method, also called Blend, overlaid the trigger image on the clean image to produce poisoned image. Nonetheless, the poisoned images created by the above methods are highly conspicuous and can be easily identified by human observers. Consequently, recent research has focused on developing more stealthy attack methods to avoid detection\cite{ning2021invisible,xue2023compression,wu2022just}. Li et al.\cite{li2021invisible} presented an invisible backdoor attack based on an autoencoder structure and Nguyen et al.\cite{nguyen2021wanet} introduced Wanet using warping-based triggers. Nevertheless, upon careful observation, it is evident that the poisoned images generated by these attacks still exhibit unnatural characteristics, undermining their invisibility. Besides, some methods concentrate on revealing potential threats of backdoor attack in the physical world\cite{xue2022ptb,li2021backdoor,xue2021robust}. Zhao et al.\cite{zhao2022natural} conducted a backdoor attack by simulating raindrop effects in real-world settings, and Emily et al.\cite{wenger2021backdoor} constructed a polluted face recognition dataset using physical objects such as glasses and stickers for their attack.

\subsection{Backdoor Defense}
The emergence of backdoor attack has prompted a growing interest in backdoor defense research to alleviate their threats. Most existing approaches primarily focus on defending against poisoning-based backdoor attacks and can be categorized into two types: defensive and detection approach. Defensive approaches aim to neutralize the negative influences of potential backdoor attack through various techniques, such as model reconstruction\cite{fan2022defense}, neuron pruning\cite{wu2021adversarial}, and knowledge distillation\cite{li2021protecting}. For example, Zheng et al.\cite{zheng2022data} propose a novel defense method called CLP, which can detect potential backdoor channels in a data-free manner and perform simple pruning on the infected DNN to repair it. ARGD\cite{xia2022eliminating} is an effective backdoor defense algorithm based on graph knowledge distillation, proposed by Xia et al. It has been shown to eradicate more backdoor triggers compared to NAD\cite{li2021neural}. Detection approaches are designed to analyze the model's behavior in order to detect the presence of injected backdoor triggers\cite{yang2021detecting,xia2021statistical,shen2021backdoor}. Neural Cleanse\cite{wang2019neural} generates potential triggers for each class of the model and computes anomaly indices for them. Subsequently, it determines whether the model has been implanted with a backdoor or not based on the anomaly index. In addition, Dong et al.\cite{dong2021black} attempted to identify backdoor attacks by utilizing their proposed gradient-free optimization algorithm to reverse potential triggers for each class.

Backdoor attack and backdoor defense can be seen as two opposing forces that motivate each other. In this paper, we compare our approach to well-known backdoor attacks and demonstrate that our attack is capable of resisting backdoor defenses.
\begin{figure}[b]
\includegraphics[width=\columnwidth]{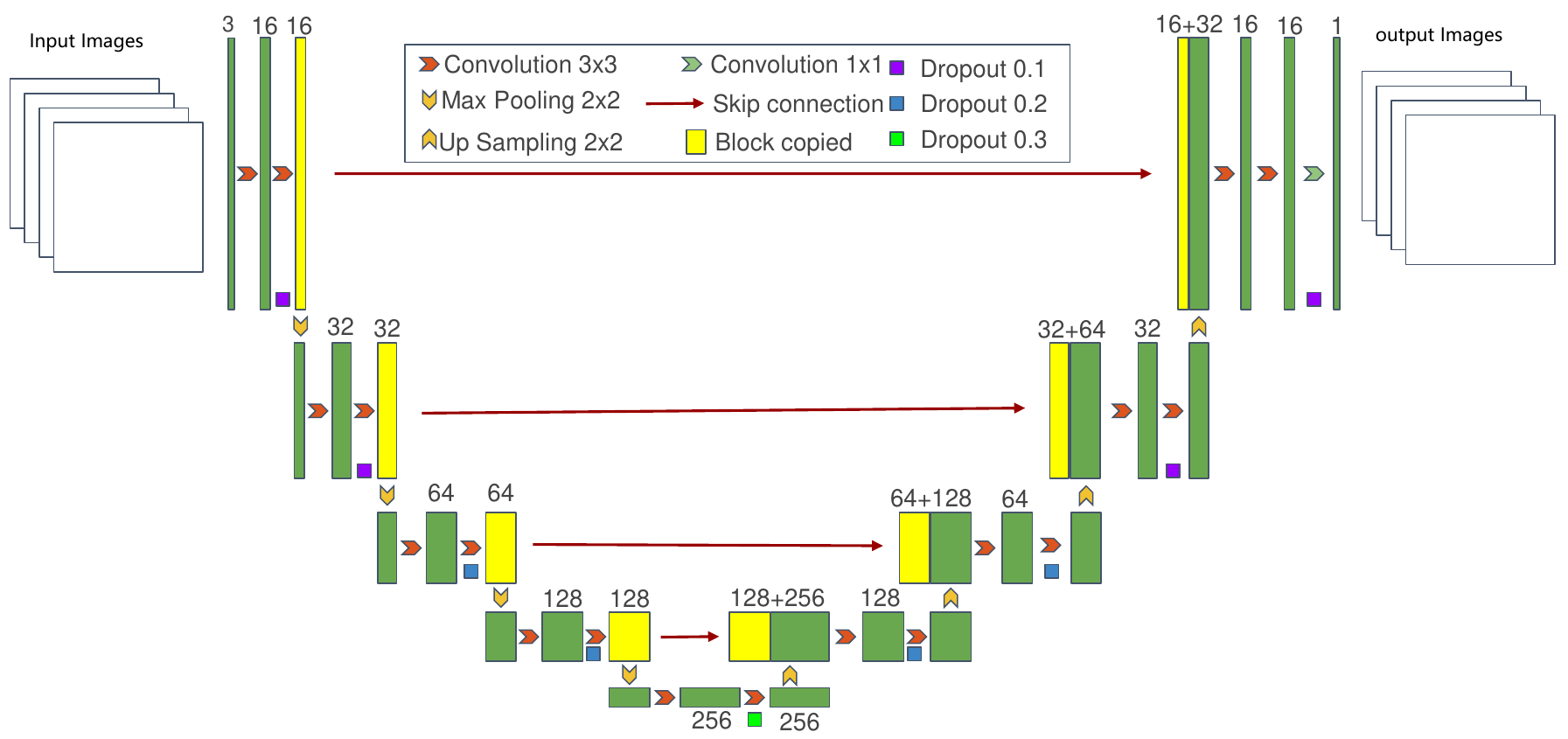}
\caption{Network structure of the U-net.} \label{fig1}
\end{figure}
\subsection{U-net}
The U-net\cite{ronneberger2015u} was originally introduced to address medical image segmentation challenges. It utilizes an U-shaped network architecture to capture both image context and location information, achieving remarkable success in the ISBI Cell Tracking Challenge 2015. Fig. \ref{fig1} illustrates the U-net's structure, which consists of an encoder and decoder. The encoder performs feature extraction and reduces image dimension through convolution and downsampling, while the decoder part upsampling to recover the features and dimension of image. 

During the entire architecture, the U-net incorporates skip-connection for feature fusion, enabling the network to get feature information from the lower dimensional convolution layers. This helps compensate for the feature loss caused by the convolutional operation. Drawing inspiration from this advantage, we employ U-net into our trigger injection network to address the issue of representation loss that arises during the generation of poisoned images.

\subsection{Spatial Attention}
\begin{figure}[htbp]
\includegraphics[width=\columnwidth]{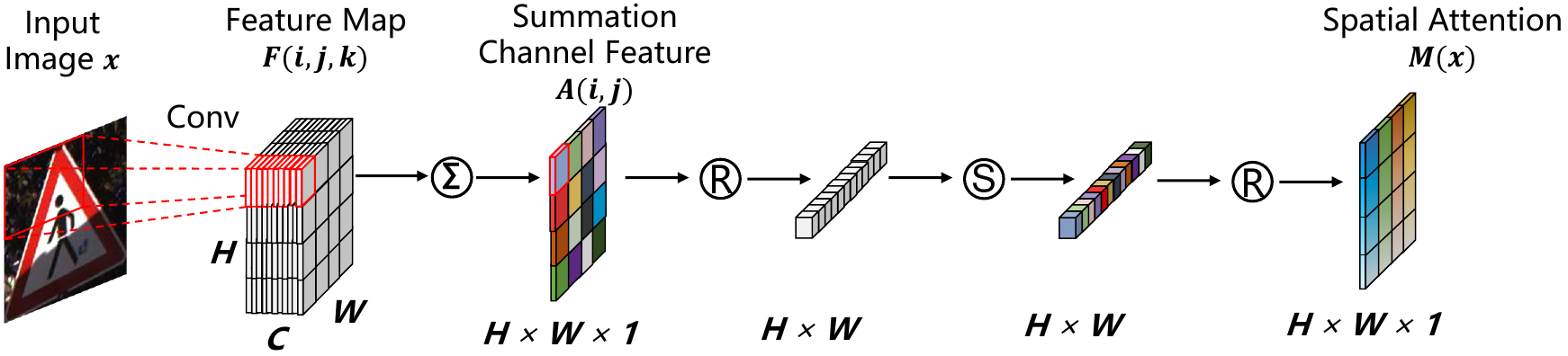}
\caption{Acquisition process of spatial attention. In this figure, ${\sum}$ means the sum operation across the channel dimension, $R$ denotes the Reshape operation, and $S$ indicates the Softmax operation. The final result is the spatial attention, which represents the important region of a given victim model for a specific image.} \label{fig2}
\end{figure}

Spatial attention refers to the ability of a model to selectively focus on specific regions of an image. One of the earliest works on spatial attention was the Recurrent Attention Model (RAM)\cite{mnih2014recurrent}, which used recurrent neural networks (RNNs) and Reinforcement Learning (RL) to learn where to attend. As depicted in Fig. \ref{fig2}, when a sample image ${x}$ is fed through a convolutional layer of the model, it produces the image feature ${F(i, j, k)\in R^{H\times W\times C}}$. Subsequently, the feature activation maps are summed across the channel dimension to obtain $A\left(i, j\right)$. This can be mathematically formalized by the following equation:
\begin{equation}
    A\left(i, j\right)=\sum_{k=1}^{C}F\left(i, j, k\right)=\sum_{k=1}^{C}{f^{h\times w}\left(x\right)},
\end{equation}
where ${f^{h\times w}}$ denotes the convolution operation with a convolution kernel of size ${h\times w}$, which varies according to the specific model architecture.

Then, the spatial attention ${M(x)\in R^{H\times W\times1}}$ is obtained through subsequent Reshape, Softmax, and Reshape operations. The entire process can be expressed as follows:
\begin{equation}
\begin{aligned}
M\left(x\right)&=Reshape\left(Softmax\left(Reshape\left(A\left(i,j\right)\right)\right)\right)\label{eq2}
\end{aligned}
\end{equation}

Spatial attention can be exploited to locate the important image regions for a given target model. By producing and embedding backdoor trigger into these regions, we can theoretically enhance the performance and robustness of the attack. This is because different images may have different attention regions for different target models, and therefore the trigger pattern can be more flexible, i.e., the trigger can vary depending on the sample and the target model.

\section{METHOD}
In this section, we will begin by explaining the concept of backdoor attack and introducing our threat model. Subsequently, we will present our novel approach for conducting backdoor attack, which utilizes spatial attention. Finally, we will provide an overview of the SATBA attack.
\subsection{Problem Formulation}
The primary objective of this research is to investigate the efficacy of backdoor attack on image classification neural networks. We define ${\mathcal{X}}$ as the input image domain and ${\mathcal{C}}$ as the corresponding ground-truth label set of ${\mathcal{X}}$. A deep neural network ${F_\theta}$ is trained to maps the input space to the label space using the dataset ${D=\{\left(x_i,y_i\right)\}}$ where $x_i\in\mathcal{X},y_i\in\mathcal{C}$ and $i=1,\ 2,\ \ldots,\ N$. For backdoor attack, the attacker carefully selects clean images from ${D}$ and generates poisoned samples. A DNN model ${{\hat{F}}_\theta}$ then is trained with a poisoned dataset ${D_p}$, which consists of a backdoored version of the ${D(i.e., D_b)}$ and remaining clean sample ${D_c}$, i.e.,
\begin{equation}
    D_p=D_b\cup D_c,
\end{equation}
accordingly, the poisoning rate is defined as ${\eta=\frac{\left|D_b\right|}{\left|D\right|}}$. As a consequence, the infected model ${{\hat{F}}_\theta}$ works as expected on the benign dataset ${D_c}$ but outputs malicious predictions when triggered by a poisoned image, that is,
\begin{equation}
\hat{F}_\theta(x_i)=y_i,\hat{F}_\theta(\hat{x_i})=y_t,
\end{equation}
where ${x_i \in D_c, y_i \in \mathcal{C}, \hat{x_i} \in D_b, y_i \not= y_t}$. The ground-truth label of $x_i$ and the attack’s target label are ${y_i}$ and ${y_{t}}$, respectively.
\begin{figure*}[ht]
    \centering
    \includegraphics[width=\textwidth]{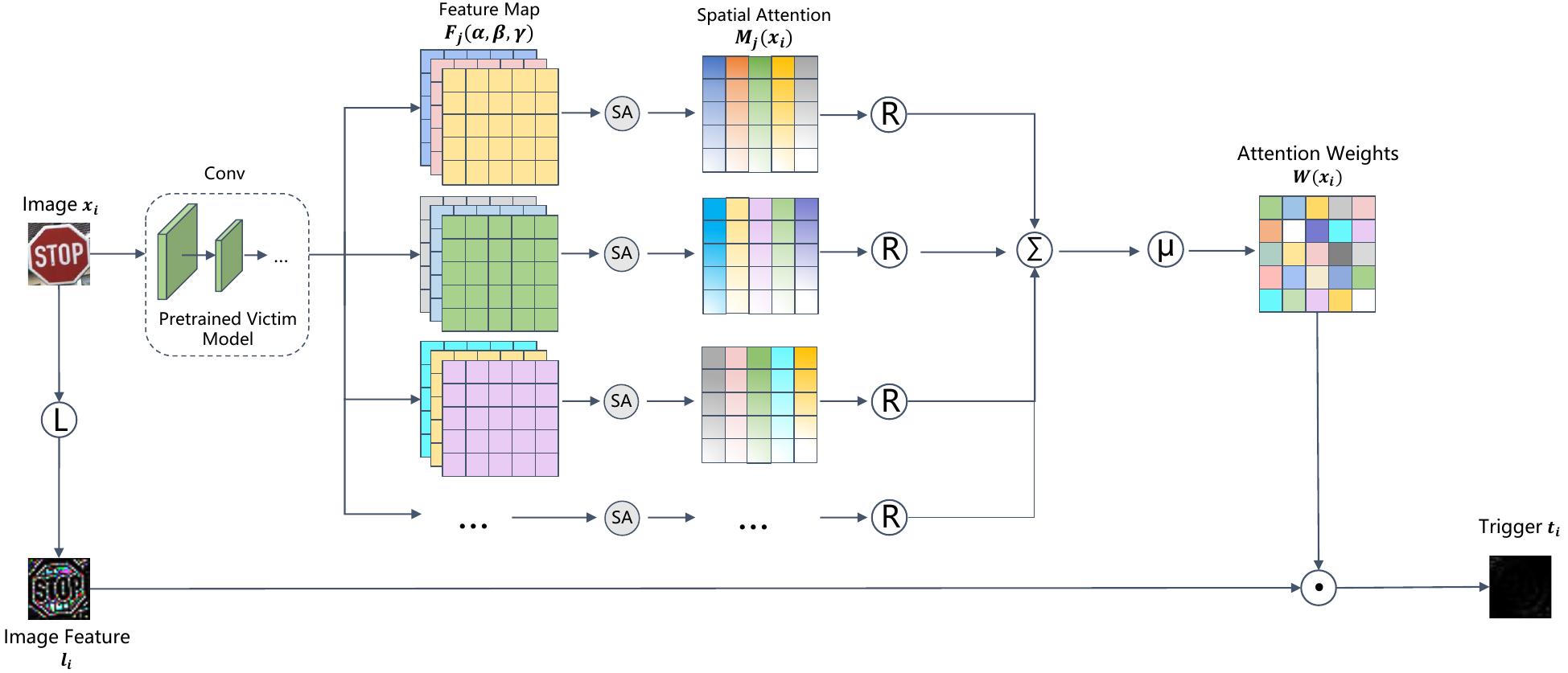}
    \caption{Our trigger generation process. It consists of three stages: (1) extracting image features from a benign sample and get the image feature ${l_i}$. (2) Feeding the clean image into a pretrained model and obtaining the spatial attention weights ${W(x_i)}$. (3) Multiplying ${W(x_i)}$ with ${l_i}$ and generate the corresponding trigger pattern ${t_i}$. It is notice that ${SA}$ denotes the calculation of spatial attention represented in Fig. \ref{fig2}.}
    \label{fig3}
\end{figure*}
\subsection{Threat Model}
\subsubsection{Attacker's Capacities}
Backdoor attack can take place at various stages of deep learning model application, including model training and deployment. Among these, attacks during the model training phase are the most prevalent and concerning. In this paper, we assume that the attacker has access to a portion of the training set and can manipulate it to inject triggers. Additionally, the attacker has knowledge of the target model's architecture, but they are unable to control the training process. Specifically, they cannot access or modify any hyperparameter settings during the training phase. After completing the model training, the attacker can only query the model but cannot change its weights.
\subsubsection{Attacker's Goals}
The primary goal of the attacker is to ensure the effectiveness of the attack, which means that a sample with the trigger can successfully activate the backdoor in the model, leading to a prediction of the target class for the polluted image. Moreover, the attacker strives to enhance the attack's stealthiness and robustness. This includes ensuring that the attack is resistant to common backdoor defenses and that the poisoned image exhibits minimal visual differences from its clean counterpart.
\subsection{Our Proposed Attack: SATBA}
\subsubsection{Trigger Generation}
Fig. \ref{fig3} provides an example of trigger generation using spatial attention. Initially, a clean image ${x_i}$ is processed through an image feature extraction algorithm ${L\left(\cdot\right)}$ (e.g., HOG\cite{pang2011efficient}, CLCM\cite{gebejes2013texture} and LPB\cite{lis2007association} etc.), resulting in image features ${l_i}$, i.e.,
\begin{equation}
    l_i=L\left(x_i\right)\label{eq5}
\end{equation}

Furthermore, $N$ feature maps ${F_j\left(\alpha,\beta,\gamma\right)}$ of ${x_i}$ are obtained by feeding ${x_i}$ into a pretrained clean model that shares the same architecture as the victim model, and the corresponding spatial attention ${M_j\left(x_i\right)}$ is computed using Eq. (\ref{eq2}) presented in Section II. Since the dimensions and sizes of spatial attentions are different from each other, we use the Reshape operation to keep them the same size as the original image. Subsequently, the weights of each ${M_j\left(x_i\right)}$ are combined using the Mean operation (represented by ${\mu}$ in Fig. \ref{fig3}) to obtain the spatial attention weights ${W\left(x_i\right)}$. Consequently, we can model this process as:
\begin{equation}
    W\left(x_i\right)=\frac{1}{N}\sum_{j=1}^{N}Reshape\left(M_j\left(x_i\right)\right).\label{eq6}
\end{equation}
Finally, we compute the dot product between the image feature ${l_i}$ and ${W(x_i)}$ yielding the trigger ${t_i}$:
\begin{equation}
    t_i=l_i\odot W\left(x_i\right).
\end{equation}

\subsubsection{Injection Architecture}
\begin{figure}[ht]
    \centering
    \includegraphics[width=\columnwidth]{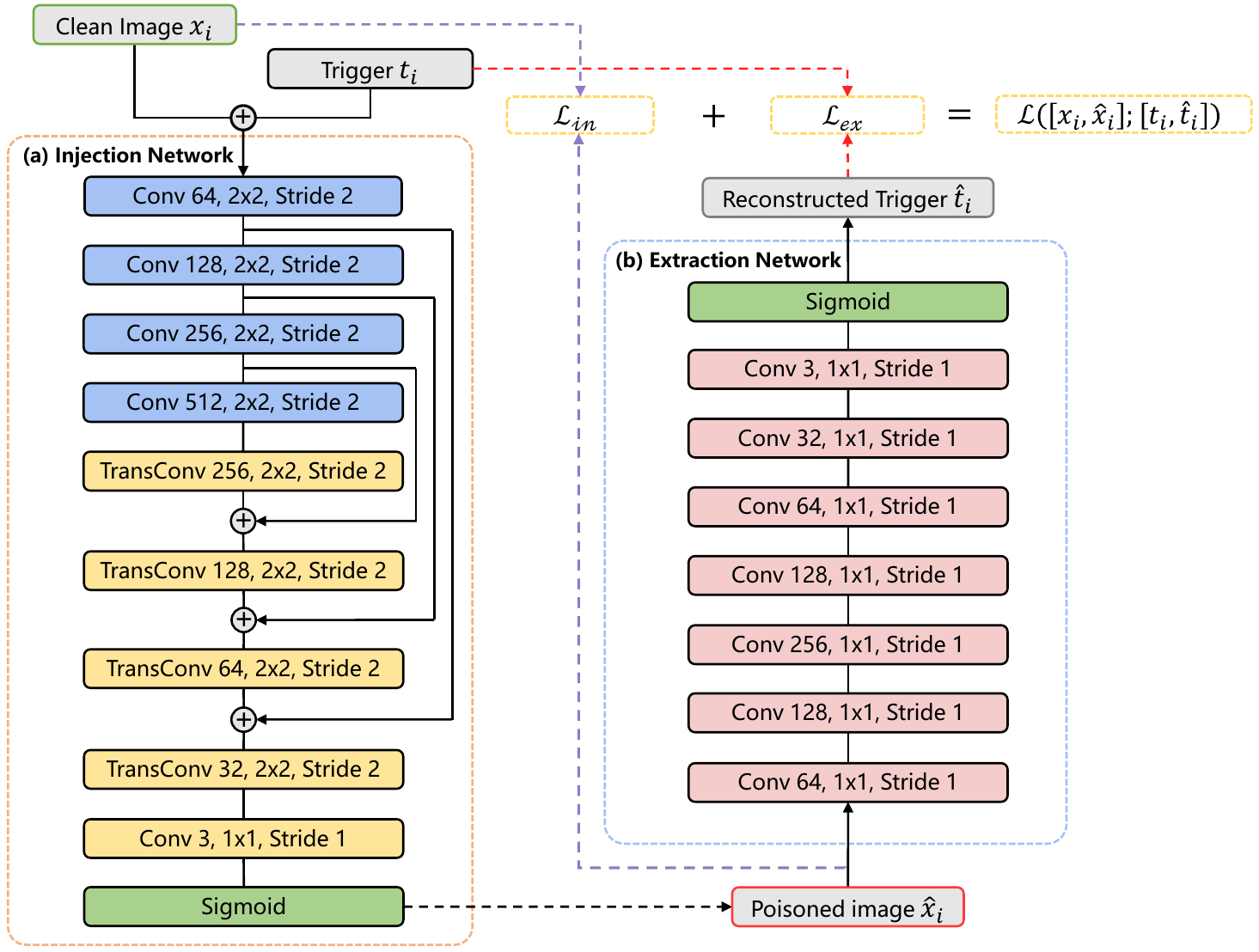}
    \caption{Our trigger injection architecture. (a) An U-net based injection network which is used to plant triggers into clean images. (b) A fully convoluted extraction network to restore trigger from poisoned images.}
    \label{fig4}
\end{figure}

Once the triggers are generated, a neural network model based on U-net is developed to conceal these triggers into clean images. As depicted in Fig. \ref{fig4}, the complete model architecture consists of two parts: the injection network and the extraction network. The primary objective of the injection network is to efficiently embed triggers into clean images, reducing the gap between poisoned image and clean image. Conversely, the extraction network is designed to recover the triggers from poisoned image, ensuring the preservation of trigger features. Upon completing the training of the entire model, we only utilize the injection network to hide triggers into benign images, while discarding the extraction network.

\textbf{Injection Network.} We concatenate the clean image ${x_i}$ with its corresponding trigger pattern ${t_i}$ and pass them to injection network to obtain the poisoned image ${{\hat{x}}_i}$, i.e.,
\begin{equation}
    {\hat{x}}_i=I_\theta\left(x_i\oplus t_i\right),
\end{equation}
where ${I_\theta\left(\cdot\right)}$ refers to the injection network, and ${\oplus}$ represents the concatenate operation.

We aim to achieve invisible hiding of the trigger pattern in the clean image by optimizing an injection loss ${\mathcal{L}_{in}}$ between the clean and poisoned image. Formally, ${\mathcal{L}_{in}}$ can be defined as follows:
\begin{equation}
    \mathcal{L}_{in}=\frac{1}{N}\sum_{i=1}^{N}\left|\left|x_i-\hat{x_i}\right|\right|^{k}.
\end{equation}

\textbf{Extraction Network.} In order to maintain the stealthiness of the backdoored image and minimize the feature loss of trigger, we employ an extraction network to recover the trigger pattern from the poisoned image obtained from the injection network:
\begin{equation}
    {\hat{t}}_i=E_\theta\left({\hat{x}}_i\right).
\end{equation}
Here, ${E_\theta\left(\cdot\right)}$ represents the extraction network and ${{\hat{t}}_i}$ denotes the reconstructed trigger image.

Besides, we can introduce an extraction loss ${\mathcal{L}_{ex}}$ to maintain the trigger features during the injection and extraction process:
\begin{equation}
    \mathcal{L}_{ex}=\frac{1}{N}\sum_{i=1}^{N}\left|\left|t_i-\hat{t_i}\right|\right|^{k}.
\end{equation}

Finally, the overall loss function is given as:
\begin{equation}
\begin{aligned}
    \mathcal{L}\left(\left[x_i,{\hat{x}}_i\right];\left[t_i,{\hat{t}}_i\right]\right) = \lambda_1\mathcal{L}_{in}+\lambda_2\mathcal{L}_{ex} \\
    = \frac{1}{N}\sum_{i=1}^{N}\left|\left|x_i-\hat{x_i}\right|\right|^{k}+\frac{1}{N}\sum_{i=1}^{N}\left|\left|t_i-\hat{t_i}\right|\right|^{k}.
\end{aligned}
\end{equation}
Empirically, we use ${L1}$ loss and set ${k=1}$. The hyper-parameters ${\lambda_1}$ and ${\lambda_2}$ are used to control the contribution of each loss term. Optimizing the above loss function allows us to efficiently minimize the feature loss of image and trigger.

\subsection{Attack Pipeline and Algorithm}
\begin{figure}
    \centering
    \includegraphics[width=\columnwidth]{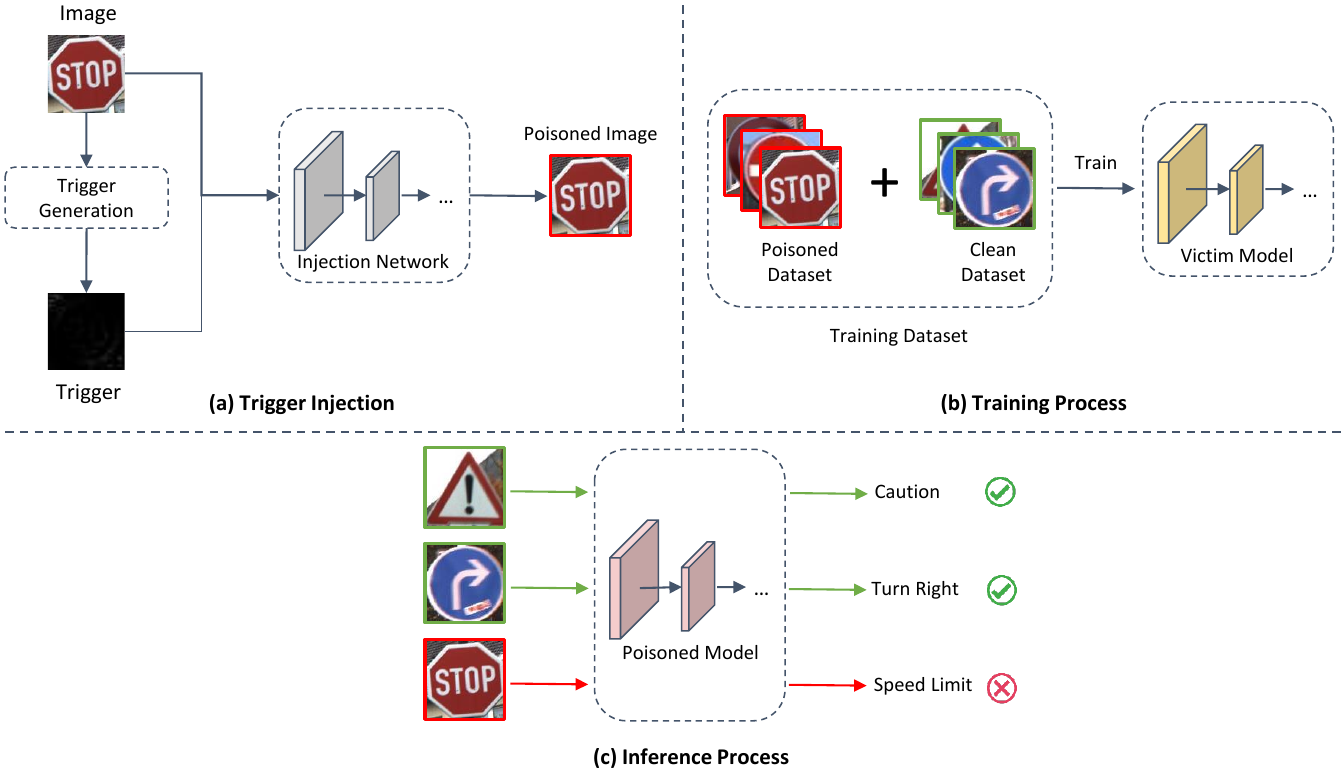}
    \caption{Overview of our proposed SATBA attack. (a) Extract image features and craft the poisoned images using the injection network. (b) Release the poisoned dataset and train the victim model. (c) The backdoor is successfully inserted into the target model.}
    \label{fig5}
\end{figure}

Fig. \ref{fig5} and Alg. \ref{alg1} illustrates the pipeline of our attack. First, the trigger generation module takes clean images as an input and produces triggers related to them. The injection network we trained then produces poisoned images by adding the triggers to a specific location of the clean images. After that, we train a victim deep model on the poisoned dataset that contains polluted images from the previous process. This leads to the successful injection of the backdoor into the target model when the training process ends.

\begin{algorithm}
    \caption{SATBA Backdoor Attack}              
    \label{alg1}
\begin{algorithmic}[1]
\REQUIRE \;
\begin{enumerate}
    \item Training set ${D=\{\left(x_i,y_i\right), i=1,\ 2,\ \ldots,\ N}\}$
    \item Target label ${y_t}$
    \item Poisoning rate ${\eta}$
    \item Victim model structure ${\mathcal{F}}$
\end{enumerate}

\ENSURE Backdoored model ${{\hat{F}}_\theta}$ \\
\STATE\textbf{// Stage I: Initialization}
\STATE Train a benign DNN ${F_\theta}$ with ${D}$ using structure ${\mathcal{F}}$
\STATE Random sample ${M=\left\lfloor\eta\ \times\ N\right\rfloor}$ data from ${D}$
\STATE \textbf{// Stage II: Trigger Generation}
\STATE Let ${D_b^\prime\gets\{\}}$
\FOR{${i = 1}$ to ${M}$}
    \STATE Feed ${x_i}$ to ${F_\theta}$ and obtain its feature maps ${F_j(\alpha,\beta,\gamma)}$
    \STATE ${l_i\gets}$ Getting image feature ${l_i}$ using Eq. (\ref{eq5})
    \STATE ${W(x_i)\gets}$ Compute spatial attention weights ${W(x_i)}$ by Eq. (\ref{eq6})
    \STATE ${t_i\gets l_i\bigodot W(x_i)}$
    \STATE ${D_b^\prime\gets D_b^\prime\cup\{\left(x_i,t_i\right)\}}$
    \STATE ${D\gets D\backslash\{\left(x_i, y_i\right)\}}$
\ENDFOR
\STATE \textbf{// Stage III: Trigger Injection}
\STATE Let ${D_b\gets\{\}}$
\STATE Train an injection architecture using ${D_b^\prime}$
\STATE Discard extraction part and obtain injection network ${I_\theta}$
\FOR{${j = 1}$ to ${M}$}
    \STATE ${{\hat{x}}_j\gets I_\theta\left(x_j\oplus t_j\right)}$
    \STATE ${D_b\gets D_b\cup\{\left({\hat{x}}_j,y_t\right)\}}$
\ENDFOR
\STATE \textbf{// Stage IV: Victim Model Training}
\STATE Let ${D_p\gets D_b\cup{D}}$
\STATE Train poisoned model of structure ${\mathcal{F}}$ on dataset ${D_p}$ and get poisoned model ${{\hat{F}}_\theta}$
\end{algorithmic}
\end{algorithm}

\section{EXPERIMENTS}
The experimental setup is explained in this section, followed by the assessment of the effectiveness of our attack. We also analyze the robustness and invisibility, and poison rate impact of SATBA.
\subsection{Experiment Setup}
\subsubsection{Dataset and Training Configuration}
In our experiments, we evaluated the performance of our attack on three standard datasets: MNIST, CIFAR10, and GTSRB using three popular deep models: AlexNet\cite{krizhevsky2012imagenet}, VGG16\cite{simonyan2014very}, and Resnet18\cite{he2016deep}. To generate the poisoned datasets $D_p$, we randomly selected image samples from each class for the three datasets using a poison rate of $\eta=0.1$. The clean images were then replaced with their corresponding poisoned samples. 

During training of the victim model, we set the learning rate to 0.1 and schedule it to decrease by a factor of 0.1 every 50 epochs, using SGD\cite{ruder2016overview} optimizer on 150 epochs. To achieve a balance between the injection network and extraction network, we found that the poisoned image and the trigger image perform well if we set $\lambda_1$ and $\lambda_2$ to 0.5 and 1.0 respectively. The trigger injection and extraction networks are trained using the Adam\cite{kingma2014adam} optimizer for 150 epochs with a learning rate of 0.001, which is gradually decreased by a factor of 0.5 when the validation loss of the model has not reduced in the previous 3 epochs.
\subsubsection{Evaluation Indicator}
The performance of our attack is evaluated using Attack Success Rate (ASR) and Clean Data Accuracy (CDA). ASR measures the effectiveness of test samples with triggers that are successfully predicted to be the target label $y_t$, while CDA indicates the accuracy of the infected model on the clean test dataset, i.e.,
\begin{equation}
\begin{aligned}
    ASR=\frac{1}{N}\sum_{i=1}^{N}\mathbb{I}\left(\hat{F}\left({\hat{x}}_i\right)=y_t\right),\\
    CDA=\frac{1}{N}\sum_{i=1}^{N}\mathbb{I}\left(\hat{F}\left(x_i\right)=y_i\right),
\end{aligned}
\end{equation}
where $\mathbb{I}\left(\cdot\right)$ is the indicator function.

A key factor in the stealthiness of a backdoor attack is the level of similarity between clean and poisoned images. Higher similarity leads to a more inconspicuous attack. To quantitatively evaluate the similarity between the clean image and the poisoned image generated by different attacks, we measure the Peak Signal-to-Noise Ratio (PSNR)\cite{hore2010image}, Structural SIMilarity index (SSIM)\cite{ye2017robust}, Mean Square Error (MSE), and Learned Perceptual Image Patch Similarity (LPIPS)\cite{zhang2018unreasonable}. 

PSNR is one of the most common and widely used objective evaluation metrics for images, which evaluates image similarity by calculating the gap between the corresponding pixels of two images using the following formulation:
\begin{equation}
\begin{aligned}
    PSNR&=10{log}_{10}\left(\frac{\left(2^n-1\right)^2}{MSE}\right),\\
    MSE&=\frac{1}{H\times W}\sum_{i=1}^{H}\sum_{j=1}^{W}\left(x\left(i,j\right)-\hat{x}\left(i,j\right)\right)^2,
\end{aligned}
\end{equation}
where $x\left(i,j\right)$ and $\hat{x}\left(i,j\right)$ represent their pixel values respectively, $H$ and $W$ are the height and width of the image, and $n$ is the number of bits of each sample value. PSNR is a unit of measurement in dB, A larger PSNR value corresponds to a smaller MSE, indicating that the two images are closer to each other with minimal distortion.

SSIM measures the similarity between two images in terms of three key image features: Luminance, Contrast and Structure. These values are carried out using the following formulas:
\begin{equation}
    \begin{aligned}
    l\left(x,\hat{x}\right)&=\frac{2\mu_x\mu_{\hat{x}}+c_1}{\mu_x^2+\mu_{\hat{x}}^2+c_1},\\
    c\left(x,\hat{x}\right)&=\frac{2\sigma_{x\hat{x}}+c_2}{\sigma_x^2+\sigma_{\hat{x}}^2+c_2},\\
    s\left(x,\hat{x}\right)&=\frac{2\sigma_{x\hat{x}}+c_3}{\sigma_x\sigma_{\hat{x}}+c_3},
    \end{aligned}
\end{equation}
where $\mu_x$ and $\mu_{\hat{x}}$ denote the mean of $x$ and $\hat{x}$, respectively. Additionally, $\sigma_x$ and $\sigma_{\hat{x}}$ represent the variance of x and $\hat{x}$, while $\sigma_{x\hat{x}}$ indicates the covariance of $x$ and $\hat{x}$. The constants $c_1$, $c_2$, and $c_3$ are determined through empirical calculations. Then, the SSIM value is calculated using the following equation:
\begin{equation}
    SSIM\left(x,\hat{x}\right)=l\left(x,\hat{x}\right)\cdot c\left(x,\hat{x}\right)\cdot s\left(x,\hat{x}\right)
\end{equation}
The SSIM value range from 0 to 1, with a higher value indicating a greater similarity between the images.

Unlike traditional similarity metrics such as PSNR and SSIM, LPIPS is a perceptual similarity metric based on deep learning. It can effectively measure the similarity between clean and poisoned image from a human perspective. A smaller LPIPS value means a higher level of similarity between the images. In this paper, we adopt LPIPS based on features learned by a pretrained AlexNet, adhering to the settings provided in the original paper of LPIPS.

\subsection{Attack Performance}
We evaluate the effectiveness of our proposed SATBA attack by comparing it with four conventional backdoor attacks, namely Badnets\cite{gu2019badnets}, Blend\cite{chen2017targeted}, Refool\cite{liu2020reflection}, and Wanet\cite{nguyen2021wanet}. To ensure a fair evaluation, we adopt an All-to-One attack strategy in which all poisoned images are labeled with the same target label $y_t$ (class 0). The results are presented in Table \ref{tab1}, which includes the ASR and CDA of different backdoor attacks on the three standard image classification datasets and DNNs. Our SATBA attack successfully poisons deep models by injecting only a small portion of the training set and achieves comparable ASR to other backdoor attacks. Meanwhile, the SATBA does not cause a substantial decrease in the validation accuracy of the infected model on clean datasets, and even shows improvement in some cases. While SATBA's ASR and CDA may not always significantly exceed those of other attacks, it is sufficient to conduct a backdoor attack against the victim DNN.
\begin{table}[ht]
    \caption{Comparison of ASR and CDA of SATBA with other attacks on different models and datasets. The "Clean" row shows the accuracy of the original model on the clean dataset. The best results are in bold, and $\ast$ indicates the same score as the best result. The second-best result is underlined.}\label{tab1}
    \resizebox{\linewidth}{!}{
    \begin{tabular}{*{8}{c}}
    \Xhline{2pt}
    Dataset→ & \multicolumn{2}{c}{MNIST} & \multicolumn{2}{c}{CIFAR10} & \multicolumn{2}{c}{GTSRB}\\
    \cmidrule(lr){2-3}\cmidrule(lr){4-5}\cmidrule(lr){6-7}
    Attack↓ & CDA & ASR & CDA & ASR & CDA & ASR \\
    \Xhline{2pt}
    Badnets AlexNet & 0.993 & 0.999 & 0.869 & 0.943 & 0.957 & 0.994 \\
    \quad \quad \quad \quad VGG16 & 0.994 & 1.000 & 0.886 & 0.967 & 0.963 & 0.995 \\
    \quad \quad \quad \quad Resnet18 & \textbf{0.996} & 1.000 & 0.898 & 0.961 & 0.962 & 0.997 \\
    \hline
    Blend \quad AlexNet & 0.992 & 1.000 & 0.885 & 0.996 & 0.960 & \textbf{0.999} \\
    \quad \quad \quad \quad VGG16 & 0.992 & 1.000 & 0.895 & 0.997 & 0.962 & 0.999 \\
    \quad \quad \quad \quad Resnet18 & 0.994 & 1.000 & 0.906 & 0.998 & 0.961 & \textbf{0.999} \\
    \hline
    Clean \quad AlexNet & 0.992 & ------ & 0.889 & ------ & 0.961 & ------ \\
    \quad \quad \quad \quad VGG16 & 0.901 & ------ & \textbf{0.902} & ------ & 0.962 & ------ \\
    \quad \quad \quad \quad Resnet18 & 0.994 & ------ & \textbf{0.911} & ------ & \textbf{0.967} & ------ \\
    \hline
    Refool \quad AlexNet & 0.992 & 1.000 & 0.879 & 0.938 & 0.959 & 0.992 \\
    \quad \quad \quad \quad VGG16 & 0.991 & 1.000 & 0.892 & 0.953 & 0.963 & 0.992 \\
    \quad \quad \quad \quad Resnet18 & 0.994 & 1.000 & 0.901 & 0.954 & \textbf{0.961} & 0.995 \\
    \hline
    Wanet \quad AlexNet & 0.994 & 0.999 & 0.886 & 0.998 & 0.961 & 0.999 \\
    \quad \quad \quad \quad VGG16 & 0.995 & 1.000 & 0.893 & 0.999 & 0.965 & 0.999 \\
    \quad \quad \quad \quad Resnet18 & 0.995 & 1.000 & 0.903 & 0.999 & 0.960 & 0.999 \\
    \hline
    SATBA(Ours) AlexNet & \textbf{0.996} & 1.000$\ast$ & \textbf{0.892} & \textbf{0.998} & 0.958 & \underline{0.996} \\
    \quad \quad \quad \quad VGG16 & \textbf{0.995} & 1.000$\ast$ & 0.888 & \textbf{0.998} & \textbf{0.965} & 0.999$\ast$ \\
    \quad \quad \quad \quad Resnet18 & 0.996$\ast$ & 1.000$\ast$ & 0.904 & \textbf{0.999} & 0.955 & 0.994 \\
    \Xhline{2pt}
    \end{tabular}}
    \label{tab:my_label}
\end{table}

\subsection{Defense Resistance}
\subsubsection{Resistance to Neural Cleanse}
\begin{figure}[ht]
    \centering
    \includegraphics[width=\columnwidth]{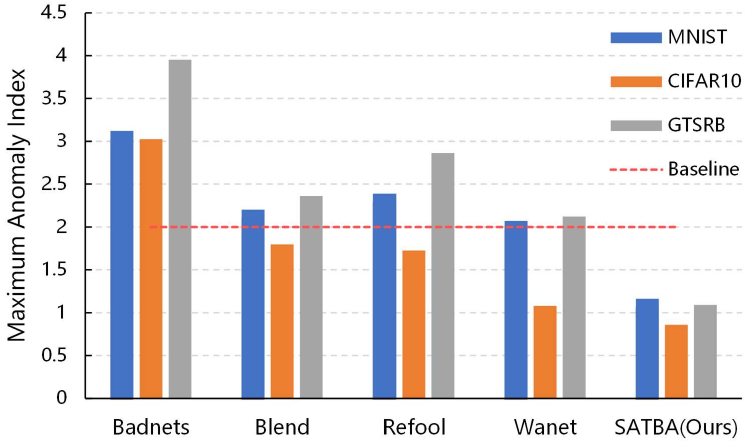}
    \caption{Detection result of Neural Cleanse for suspicious Resnet18 model trained with different backdoor attacks on MNIST, CIFAR10 and GTSRB.}
    \label{fig6}
\end{figure}

We assessed the effectiveness of our proposed SATBA attack against backdoor defense using the Neural Cleanse (NC)\cite{wang2019neural}. NC generates potential triggers for each class of the model being tested and calculates an Anomaly Index for them, with a higher Maximum Anomaly Index(MAD) indicating a greater likelihood of a backdoor being embedded in the DNN. When the MAD is greater than 2, NC considers a deep model to contain a backdoor. As depicted in Fig. \ref{fig6}, NC was unable to detect the backdoor model injected by SATBA, confirming its ability to evade backdoor defense. Furthermore, our infected model had a lower MAD compared to other DNNs trained using common backdoor attacks, indicating that SATBA has greater resistance to backdoor defense.
\subsubsection{Resistance to AEVA}
\begin{figure}[ht]
    \centering
    \includegraphics[width=\columnwidth]{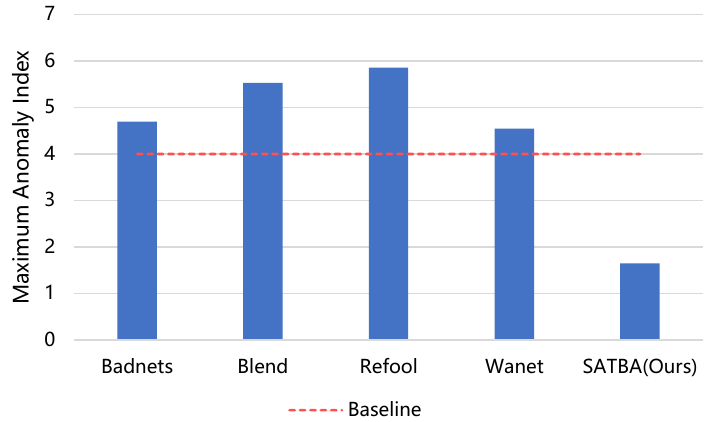}
    \caption{AEVA Maximum Anomaly Index of different well-known backdoor attacks compared to our attack.}
    \label{fig7}
\end{figure}

Recently, Guo et, al.\cite{guo2021aeva} proposed a black-box hard-label backdoor defense called AEVE. Contrary to NC, which is white box defense, it calculates the Anomaly Index by performing extreme value analysis on the adversarial map obtained using the Monte-Carlo gradient estimation. AEVE identify the class whose corresponding MAD larger than 4 as infected. We conducted the detection on CIFAR-10 with Resnet18, following the experimental setup in the original AEVA code. Fig. \ref{fig7} displays results of AEVE detection for a suspicious ResNet18 model trained with various backdoor attacks on CIFAR10. The MAD for certain backdoor attacks exceed the threshold of 4 in specific categories. However, our proposed attack, SATBA, demonstrates a significantly lower MAD, well below the threshold value of 4, indicating its strong resistance to AEVA.
\subsection{Stealthiness Analyze}
\begin{figure}
    \centering
    \includegraphics[width=\columnwidth]{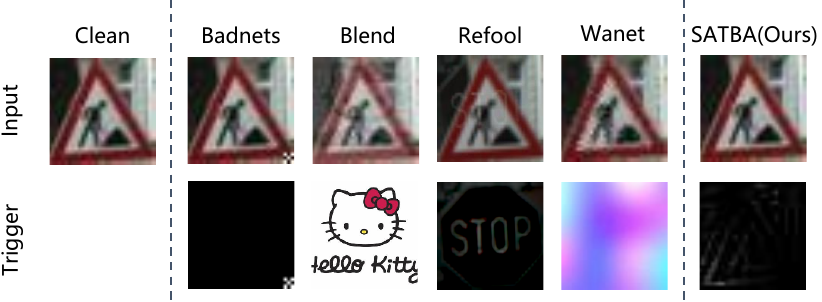}
    \caption{Visualization of the poisoned images and their corresponding triggers generated by various backdoor attacks on class 25 of the GTSRB dataset. The top row from left to right displays the original image and the backdoored image crafted by Badnets, Blend, Refool, Wanet and our proposed SATBA, while the lower row shows the corresponding triggers for each poisoned image on the upper row.}
    \label{fig8}
\end{figure}

Fig. \ref{fig8} provides a visual comparison of poisoned images and their corresponding triggers, which are generated through various backdoor attacks on the GTSRB dataset. In contrast to Badnets, Blend, Refool, and Wanet, the poisoned image created by SATBA appears more natural and closely resembles the clean image, making it less detectable by humans. Moreover, its corresponding trigger is more relevant to the clean image and is imperceptible, which is crucial for ensuring attack stealthiness.
\begin{table*}[ht]
\caption{Attack stealthiness measure by PSNR, SSIM, MSE and LPIPS compared with other attack methods.}\label{tab2}
\resizebox{\linewidth}{!}{
\begin{tabular}{*{13}{c}}
\toprule
Dataset→ & \multicolumn{4}{c}{MNIST} & \multicolumn{4}{c}{CIFAR10} & \multicolumn{4}{c}{GTSRB} \\
\cmidrule(lr){2-5}\cmidrule(lr){6-9}\cmidrule(lr){10-13}
Attack↓ & PSNR & SSIM & MSE & LPIPS & PSNR & SSIM & MSE & LPIPS & PSNR & SSIM & MSE & LPIPS \\
\hline
Badnets & 24.0935 & \textbf{0.9874} & 253.3595 & 0.0013 & 30.8597 & \textbf{0.9935} & 89.0661 & \textbf{0.0015} & 27.6426 & \textbf{0.9914} & 151.2264 & 0.0076\\
Blend   & 15.9751 & 0.5637 & 1644.4315 & 0.0215 & 20.2685 & 0.7835 & 652.2303 & 0.0313 & 18.6797 & 0.6788 & 961.7501 & 0.0796\\
Refool  & 12.4355 & 0.4612 & 6023.4371 & 0.0734 & 17.2960 & 0.6920 & 1812.7980 & 0.0668 & 14.9439 & 0.5442 & 3440.4423 & 0.1578\\
Wanet  & 23.5286 & 0.9314 & 298.6892 & 0.0090 & 29.2407 & 0.9511 & 90.2150 & 0.0077 & 32.3926 & 0.960 & 78.3503 & 0.0863\\
SATBA   & \textbf{47.2693} & \underline{0.9862} & \textbf{1.4735} & \textbf{0.0001} & \textbf{36.8021} & \underline{0.9857} & \textbf{22.8828} & \underline{0.0060} & \textbf{36.6371} & \underline{0.9802} & \textbf{21.5731} & \textbf{0.0065}\\
\bottomrule
\end{tabular}}
\end{table*}
We conducted experiments to evaluate the similarity of SATBA on MNIST, CIFAR10, and GTSRB datasets by randomly selecting 1000 images from the poisoned test set. As shown in Table \ref{tab2}, our proposed attack achieved excellent scores in all similarity metrics, including the highest PSNR and lowest MSE values for all three datasets. Although the SSIM of Badnets was better than ours, our attack was very close to the best result. In terms of LPIPS, our SATBA shows significant improvement on MNIST and GTSRB compared to others. Notably, while the LPIPS of Badnets was lower than that of SATBA, our attack achieved the second-best result and was the closest to Badnets among all attacks.

\subsection{Poisoning Rate}
\begin{figure}
\begin{minipage}[t]{\columnwidth}  
      \centering
      \label{fig:subfig5}\includegraphics[width=\columnwidth]{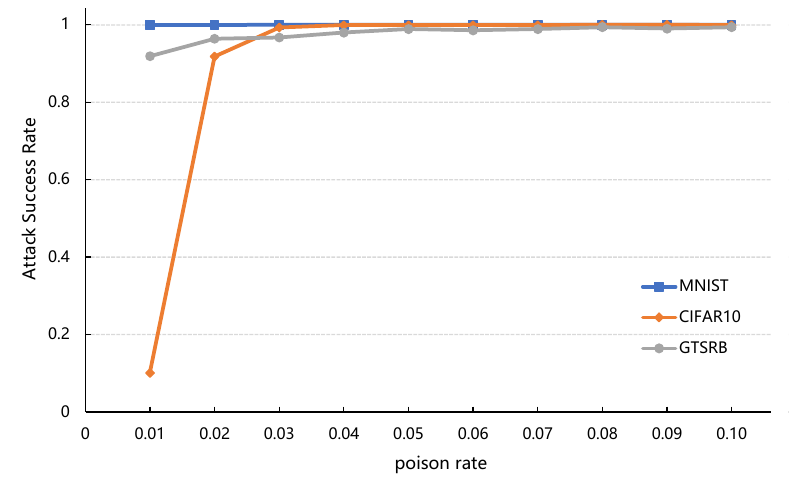}
  \end{minipage}
  \begin{minipage}[t]{\columnwidth}
      \centering
      \label{fig:subfig6}\includegraphics[width=\columnwidth]{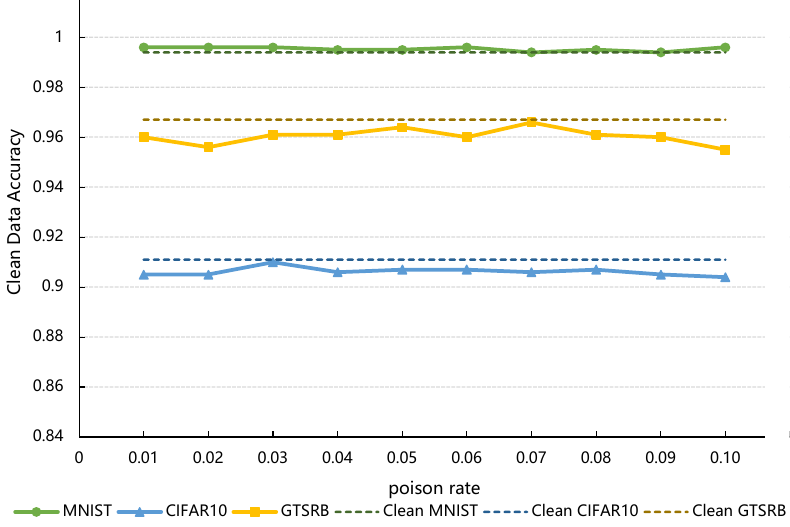}
  \end{minipage}
  \caption{The impact of poisoning rate in our attack with three datasets on Resnet18.}
  \label{fig9}
\end{figure}

To examine how the poisoning rate affects the attack success rate, we conducted experiments on Resnet18 using different datasets. The results, as presented in Fig. \ref{fig9}, demonstrate that our attack achieves a high ASR while maintaining a stable CDA on all three standard datasets. Specifically, with only 0.01 of training images poisoned, SATBA achieves nearly 100\% ASR on MNIST. For CIFAR10 and GTSRB, the proposed attack performs well with ASR greater than 0.92 when the poisoning rate is over 0.02. Additionally, the victim model's CDA remains in a normal range, with no distinguishable difference from a clean DNN (less than 0.04). This experiment validates the effectiveness of our SATBA without sacrificing the accuracy of the poisoned model on the clean dataset.
\section{DISCUSSIONS}
The spatial attention has found extensive applications in the realm of artificial intelligence technology. Our research substantiates the feasibility of employing this technique for backdoor attacks, thereby enabling real-world attackers to manipulate the target Deep Neural Network (DNN) model. Consequently, there arises a pressing need for the development of more robust and secure neural networks or the advancement of backdoor defense. 

Our forthcoming work will center on investigating the transferability of our trigger, examining whether a trigger generated from a specific dataset-DNN pair can successfully attack other models. Additionally, we intend to enhance the performance of our approach by incorporating Resnet blocks into our trigger injection network, thereby improving the feature extraction capabilities of our injection architecture. Finally, we will focus on exploring a viable defense method to mitigate this type of backdoor threat.

\section{CONCLUSIONS}
This paper presents a novel technique for creating invisible backdoor attack on deep neural networks (DNNs). Our approach involves using spatial attention to identify the focus area of a victim model on clean data enabling us to generate a unique trigger corresponding to that particular sample. An U-net based model is then employed to produce poisoned images while optimizing the feature loss of both the images and triggers. Through extensive experimentation, we have demonstrated the high effectiveness of our proposed method in generating imperceptible poisoned images that successfully attack DNNs. 

\section*{ACKNOWLEDGEMENTS}
This work was supported in part by the National Key Research and Development Program of China under Grant 2020YFB1710005 and in part by Natural Science Foundation of Shandong Province under Grant ZR2022MF299.

\section*{REFERENCES}
\bibliographystyle{ieeetr}
\bibliography{ref}

\vspace{12pt}
\end{document}